\documentclass{ijmpc}

\begin{document}

\markboth{X.-J. Xu et al.} {Mutual Selection in Network Evolution:
the Role of the Intrinsic Fitness}

\catchline{}{}{}{}{}

\title{MUTUAL SELECTION IN NETWORK EVOLUTION: THE ROLE OF THE INTRINSIC FITNESS}

\author{XIN-JIAN XU$^{*}$ and LIU-MING ZHANG}

\address{Department of Mathematics, College of Science, Shanghai University, Shanghai 200444, China
\\Institute of Systems Science, Shanghai University, Shanghai 200444, China\\$^{*}$xinjxu@shu.edu.cn}

\author{LI-JIE ZHANG}

\address{Department of Physics, College of Science, Shanghai
University, Shanghai 200444, China}

\maketitle

\begin{history}
\received{Day Month Year}
\revised{Day Month Year}
\end{history}

\begin{abstract}
We propose a new mechanism leading to scale-free networks which is
based on the presence of an intrinsic character of a vertex called
fitness. In our model, a vertex $i$ is assigned a fitness $x_i$,
drawn from a given probability distribution function $f(x)$. During
network evolution, with rate $p$ we add a vertex $j$ of fitness
$x_j$ and connect to an existing vertex $i$ of fitness $x_i$
selected preferentially to a linking probability function
$g(x_i,x_j)$ which depends on the fitnesses of the two vertices
involved and, with rate $1-p$ we create an edge between two already
existed vertices with fitnesses $x_i$ and $x_j$, with a probability
also preferential to the connection function $g(x_i,x_j)$. For the
proper choice of $g$, the resulting networks have generalized power
laws, irrespective of the fitness distribution of vertices.

\end{abstract}

\ccode{PACS Nos.: 89.75.Hc, 89.75.Fb.}

Complex networks are powerful tools to describe a large variety of
biological, social, and technical networks. A network is a
mathematical object which consists of vertices connected by edges.
Despite differences in their nature, many real-world networks are
characterized by similar topological properties, in contrast to
those obtained by traditional random graphs. One of the most
interesting phenomena is the scale-free (SF) behavior, which means a
power-law distribution of connectivity, $P(k)\sim k^{-\gamma}$,
where $P(k)$ is the probability that a vertex in the network is of
degree $k$ and $\gamma$ is a positive real number determined by the
given network. In order to understand how SF networks arise, much
work has been done in the past decade. It has been shown that growth
and preference seem to be the principal mechanisms for SF behavior.

The exploring the preference can be directed in two classes. The
first class of research is based on the \emph{rich-get-richer} rule,
which was implemented by newcomers preferential connecting to old
vertices with certain topological
characteristics.\cite{Barabasi99,Krapivsky00,Dorogovtsev00,Barrat04,Andrade05,Soares06}
In the best known Barab\'{a}si-Albert (BA) Model,\cite{Barabasi99}
the network grows at a constant rate and new vertices attach to old
ones with probability $\Pi(i) \sim k_{i}$. In this way, vertices of
high degree are more likely to receive further edges from newcomers.
In fact, this extreme assumption is not always available for many
networks when their sizes are huge. The similar mechanism was
also used in weighted networks, for instance, the network grows at a
constant rate and new vertices attach to old ones with probability
$\Pi(i) \sim s_{i}$, where $s_{i}= \sum_{j \in U(i)}w_{ij}$ is the
strength of vertex $i$ and the sum runs over the set $U(i)$ of
neighbors of $i$.\cite{Barrat04} The second class of research
utilizes the \emph{fit-get-richer} mechanism, which was carried out
by newcomers preferential connecting to old vertices with high
intrinsic
fitnesses.\cite{Bianconi01,Ergun02,Caldarelli02,Servedio04,Garlaschelli04,Bedogne06}
This is better adapted to model certain networks where topological
properties are essentially determined by \lq\lq physical\rq\rq
information intrinsically related to the role played by each vertex
in the network, such as the ability of an individual, the content of
a web page, or the innovation of a scientific article.

Caldarelli et al recently introduced a varying vertex fitness
model,\cite{Caldarelli02} where they consider an undirected graph of
$N$ vertices. At every vertex $i$ a fitness $x_i$, which is a real
number measuring its importance or rank, is assigned. Fitnesses are
random numbers taken from a given probability distribution $f(x)$.
For every couple of vertices, $i$ and $j$, an edge is created with
probability $g(x_i,x_j)$ (a symmetric function of its arguments)
depending on the \lq\lq importance\rq\rq of both vertices, i.e., on
$x_i$ and $x_j$. Actually this is a natural generalization of the
classic Erd\"{o}s-R\'{e}nyi graph.\cite{Erdos59} Although it is a
static model, the network recovers the power-law behavior of degree,
betweenness, and clustering coefficient.\cite{Caldarelli02} On the
other hand, Bedogne and Rodgers proposed a growing network with
intrinsic vertex fitnesses.\cite{Bedogne06} Besides employing the
edge-created mechanism suggested in Ref.~\refcite{Caldarelli02},
they also considered two cases of new vertices connecting to old
ones, uniform or degree-preferential. The interplay between the
fitness linking mechanism and uniform attachment results in an
exponential degree distribution for any fixed fitness $x$, while the
degree-preferential attachment instead induces that the degree
distribution decays as a power law.\cite{Bedogne06}

Models of the first class often present us such an evolution
picture: old vertices are passively attached by newcomers according
to the degree- (strength-) preferential mechanism. On the contrary,
models belongs to the second class pay much attention to the
creation and reinforcement of internal connections. Combining above
two aspects, we argue that the connection between two vertices is
the result of their mutual affinity and attachment. Not only for
interactions among new vertices and old ones, but also for that
among old vertices, which we call \lq\lq mutual selection\rq\rq.
Motivated by this, we suggest an evolving network model ruled by the
fitness-dependent selection dynamics. The generated network has a
good right-skewed distribution of degrees.

The present model starts from an initial $m$ isolated seeds and each
vertex $i$ is endowed with a fitness $x_i \ge 0$, drawn from a given
probability distribution $f(x)$. At each time step, we perform
either of the following two operations. (i) With rate $p \in (0,1)$
we add a new vertex $j$ of fitness $x_j \in f(x)$ to the network.
The new vertex connects to an existing vertex $i$ of fitness $x_i$
selected preferentially to a linking probability function
$g(x_i,x_j)$ which is symmetric and dependent on the associativity
of the both vertices. (ii) With rate $1-p$ we create an edge between
two vertices, $i$ and $j$, already presented in the network with the
probability also preferential to their integration $g(x_i,x_j)$.
After $t$ time steps, this scheme generates a network of $m+pt$
vertices and $t$ links. Notice that either process is chosen in the
network growth, only one edge is added to the system at each time
step (duplicate and self-connected edges are forbidden), however,
this is not essential.

In our model, each vertex is assigned a fitness, either initial
seeds or subsequent newcomers. Denoting $N_{k}(x,t)$ the average
number of vertices with degree $k$ and fitness $x$ at time $t$, we
can write out the rate equation for network evolution
\begin{eqnarray}
\frac{\partial N_{k}(x,t)}{\partial t} &=& \frac{p\int_{0}^{\infty}
f(x')g(x,x')[N_{k-1}(x,t)-N_{k}(x,t)]\mathrm{d}x'}{\int_{0}^{\infty}
f(x')\sum_{k=0}^{\infty}\int_{0}^{\infty}
g(x,x')N_{k}(x,t)\mathrm{d}x\mathrm{d}x'} \nonumber\\
&+& p\delta_{k,1}f(x) \nonumber\\
&+& 2(1-p)[N_{k-1}(x,t)-N_{k}(x,t)] \nonumber\\
&\times& \frac{\sum_{k=0}^{\infty}\int_{0}^{\infty}
g(x,x')N_{k}(x',t)\mathrm{d}x'}{\sum_{k=0}^{\infty}\int_{0}^{\infty}
\sum_{k=0}^{\infty}\int_{0}^{\infty}
g(x,x')N_{k}(x',t)\mathrm{d}x'N_{k}(x,t)\mathrm{d}x}. \label{rateeq}
\end{eqnarray}
The first term on the right hand side (rhs) of Eq. (\ref{rateeq})
represents the change in the average number of the vertices with
degree $k$ and fitness $x$ due to process (i). The second term on
the rhs accounts for the continuous introduction, with rate $p$, of
new vertices with fitnesses drawn from the probability distribution
$f(x)$. The last term on the rhs represents the change in the
average number of the vertices with degree $k$ and fitness $x$ due
to process (ii). We also define
\begin{equation}
N(x,t)=\sum_{k=0}^{\infty}N_{k}(x,t), \label{defnxt}
\end{equation}
and
\begin{equation}
N(t)=\int_{0}^{\infty}N(x,t)\mathrm{d}x, \label{defnt}
\end{equation}
as the average number of the vertices of fitness $x$ at time $t$ and
the average number of the vertices at time $t$, respectively.
Summing Eq. (\ref{rateeq}) over $k$ we obtain
\begin{equation}
\frac{\partial N(x,t)}{\partial t}=pf(x), \label{diffnxt}
\end{equation}
which yields
\begin{equation}
N(x,t)=pf(x)t+mf(x). \label{expressnxt1}
\end{equation}
Integrating Eq. (\ref{expressnxt1}) we find as expected $N(t)=pt+m$,
and therefore Eq. (\ref{expressnxt1}) can be rewritten as
\begin{equation}
N(x,t)=f(x)N(t). \label{expressnxt2}
\end{equation}
Now we can rewrite the integrals in the third term on the rhs of Eq.
(\ref{rateeq}) in terms of $f$ and $g$
\begin{eqnarray}
& &\frac{\sum_{k=0}^{\infty}\int_{0}^{\infty}
g(x,x')N_{k}(x',t)\mathrm{d}x'}{\sum_{k=0}^{\infty}\int_{0}^{\infty}
\sum_{k=0}^{\infty}\int_{0}^{\infty}
g(x,x')N_{k}(x',t)\mathrm{d}x'N_{k}(x,t)\mathrm{d}x} \nonumber\\
&=& \frac{1}{N(t)}\frac{\int_{0}^{\infty}
g(x,x')f(x')\mathrm{d}x'}{\int_{0}^{\infty} \int_{0}^{\infty}
g(x,x')f(x)f(x')\mathrm{d}x\mathrm{d}x'} \nonumber\\
&=& \frac{A(x)}{N(t)}, \label{rewthirdterm}
\end{eqnarray}
where
\begin{equation}
A(x)=\frac{\int_{0}^{\infty}
g(x,x')f(x')\mathrm{d}x'}{\int_{0}^{\infty}\int_{0}^{\infty}
g(x,x')f(x)f(x')\mathrm{d}x\mathrm{d}x'}. \label{defax}
\end{equation}
Furthermore, we assume that
\begin{equation}
B(t)=\int_{0}^{\infty} f(x')\sum_{k=0}^{\infty}\int_{0}^{\infty}
g(x,x')N_{k}(x,t)\mathrm{d}x\mathrm{d}x', \label{defbt}
\end{equation}
and the differential of which reads
\begin{equation}
\frac{\partial B(t)}{\partial t}
=p\int_{0}^{\infty}f(x')\int_{0}^{\infty}g(x,x')f(x)\mathrm{d}x\mathrm{d}x'
=C, \label{diffbt}
\end{equation}
where $C$ is a constant. Thus $N_{k}(x,t)$ grows linearly with $t$,
and we introduce the function $D_{k}(x)$ such that
\begin{equation}
N_{k}(x,t)=D_{k}(x)t. \label{defdkx}
\end{equation}
The degree distribution of vertices can be estimated from $D_{k}(x)$
instead. Substituting Eqs. (\ref{rewthirdterm}) and (\ref{defdkx})
into Eq. (\ref{rateeq}) gives the recursive equation for $D_{k}(x)$
\begin{eqnarray}
D_{k}(x) &=& \frac{\frac{p}{c}\int_{0}^{\infty}
f(x')g(x,x')\mathrm{d}x'+\frac{2(1-p)}{p}A(x)}{\frac{p}{c}\int_{0}^{\infty}
f(x')g(x,x')\mathrm{d}x'+\frac{2(1-p)}{p}A(x)+1}D_{k-1}(x) \nonumber\\
&+&
\frac{p\delta_{k,1}f(x)}{\frac{p}{c}\int_{0}^{\infty}f(x')g(x,x')\mathrm{d}x'+\frac{2(1-p)}{p}A(x)+1}.
\label{recursivedkx1}
\end{eqnarray}
By defining
\begin{equation}
H(x)=\frac{p}{c}\int_{0}^{\infty}f(x')g(x,x')\mathrm{d}x'+\frac{2(1-p)}{p}A(x),
\label{defhx}
\end{equation}
we can rewrite Eq. (\ref{recursivedkx1}) as
\begin{equation}
D_{k}(x)=\frac{H(x)}{H(x)+1}D_{k-1}(x)+\frac{p\delta_{k,1}f(x)}{H(x)+1},
\label{recursivedkx2}
\end{equation}
which can be solved recursively
\begin{equation}
D_{k}(x) = \frac{pf(x)H^{k-1}(x)}{[H(x)+1]^{k}}. \label{soludkx}
\end{equation}
The result demonstrates that for every fixed $x$, the degree
distribution of the generated network should follows the
right-shewed behavior. Moreover, the mutual selection rule presented
here brings on the proportionality of the vertex degree to its
fitness, which means that $H(x)$ is an implicit function of $k$.
Thus, given proper forms of the linking probability function
$g(x_i,x_j)$, one can construct networks with power-law degree
distributions.

\begin{figure}[ph]
\centerline{\psfig{file=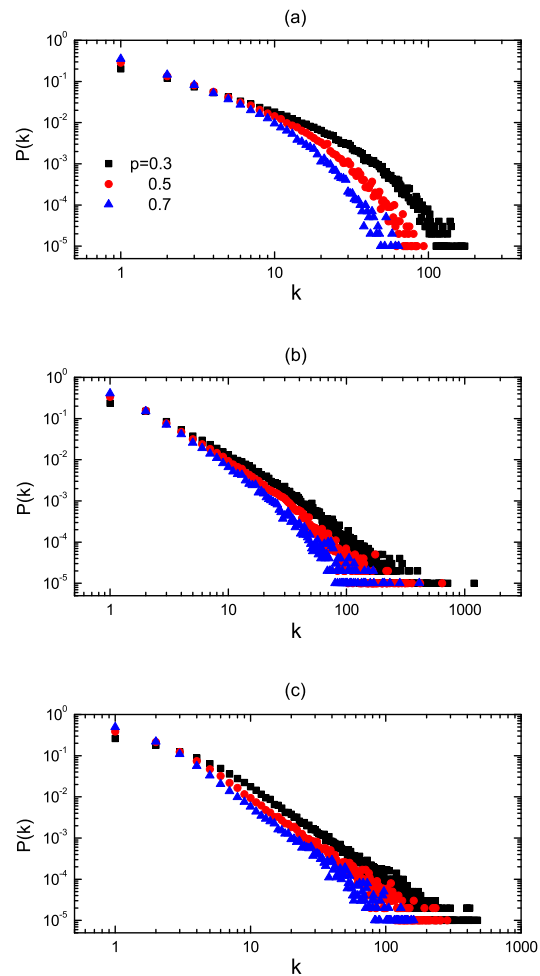,width=8cm}} \vspace*{8pt}
\caption{(color online) Degree distributions of vertices of the
generated networks for different fitness distribution functions:
uniform (a), exponential $f(x)=e^{-x}$ (b), and power-law
$f(x)=x^{-3}$ (c). The linking probability function is
$g(x_i,x_j)=x_i x_j$. Each plot corresponds to one experiment of
network generation with parameters $N=10^5$ and $m=10$.
\label{fig1}}
\end{figure}

To test above argument, we present computer simulations of the
model, as shown in Fig. \ref{fig1}. We choose the simplest case
$g(x_i,x_j)=x_i x_j$ and plot degree distributions for three kinds
distribution functions of vertex fitnesses: uniform, exponential,
and power-law. Even for this basic form of $g$, one can still notice
the generalized power laws of the degree distribution in all cases,
in agreement with analytical predictions.

In summary, we have presented an simple model to justify the
ubiquity of SF networks in nature, which results from the mutual
selection rule based on a symmetric linking probability function
$g(x_i,x_j)$ dependent on the affinity of the intrinsic fitnesses of
the involved vertices, $i$ and $j$. We found that it is always
possible to find a proper form of $g$ so that the generated network
is SF in spite of the fitness distribution. In case that the values
of vertex degrees are not available, we believe that the present
model is relatively suitable.

The authors acknowledge financial support from NSFC (No. 10805033)
and STCSM (No. 08ZR1408000). This work is sponsored by the
Innovation Foundation of Shanghai University.

\end{document}